\documentclass[12pt]{article}
\usepackage{latexsym}
\usepackage{graphics}
\newcounter{saveeqn}

\begin{document}

\begin{titlepage}\hfill HD-THEP-98-13\\[5 ex]

\begin{center}{\Large\bf Optimized $\delta $-Expansion in QCD; \\
a challenge}\\[5 ex]
{\bf Dieter Gromes}\\[3 ex] Institut f\"ur
Theoretische Physik der Universit\"at Heidelberg\\ Philosophenweg 16,
D-69120 Heidelberg \\ E - mail: d.gromes@thphys.uni-heidelberg.de
\\[10ex]
 \end{center}

{\bf Abstract: } We split the Yang-Mills Lagrangian into a free and an
interaction part in such a way, that the free part is non-local and
contains an arbitrary form factor. Manifest gauge invariance is
guaranteed by connecting the field-strength tensors at different
space-time points by a string. As a result the gluon propagator,
which, due to the presence of the string, now contains many different
contributions, comes out strictly transversal in one-loop order. We
also calculate the ghost self-energy and the ghost-gluon vertex in
one-loop order. Subsequently we discuss how one can determine the
``optimal'' form of the form factor. To apply well known principles
like ``fastest apparent convergence'' or ``principle of minimal
sensitivity'' one has first to solve some problems connected with
divergences and renormalization. Here we concentrate on the
calculation of the anomalous dimensions and the $\beta $-function.
This is technically simpler because only the divergent contributions
of the integrals have to be determined. The $\beta $-function becomes
independent of the gauge parameter as it should. A puzzle with respect
to the principle of minimal sensitivity shows up. Really interesting
non-perturbative results are expected when applying the above
principles directly to the propagator. For the $\beta $-function a two
loop calculation would be required to obtain non-trivial results.

 \vfill \centerline{March 1998}

\end{titlepage}

\section{Introduction}

The ``optimized $\delta $-expansion'' , also called ``linear $\delta
$-expansion'' or, more appropriately, ``variation perturbation
theory'', is a powerful method which combines the
merits of perturbation theory with those of variational approaches.
The underlying idea is simple. Generically, the Lagrangian is split
into a free and an interacting part in such a way that an arbitrary
parameter $\lambda $ (or more) is artificially introduced. The
interacting part is multiplied by a factor $\delta $ which serves as
formal expansion parameter and is put equal to one at the end. The
exact solution should be independent of the parameter $\lambda $
while any approximate solution will, of course, depend on it. One way
to fix the value of $\lambda $ is the ``principle of minimal
sensitivity'' (PMS): It demands that the approximate solution should
depend as little as possible on the parameter. This means that
$\lambda $ should be chosen such that the quantity to be calculated
has an extremum. In this way the result becomes non-perturbative
because $\lambda $ becomes a non-linear function of the coupling
constant. In every order of perturbation theory the optimal value of
the parameter $\lambda $ has to be calculated again.

The method is now well established and it is therefore impossible to
give all references. For the older literature we refer to Stevenson
\cite{stev85} and references therein, some more recent references can
be found, e.g. in \cite{gro}.

The field-theoretical applications have been rather modest up to now
as far as the splitting of the Lagrangian into a free and an
interacting part is concerned. Essentially only the mass parameter was
used as variational parameter in the free part of the Lagrangian. The
method is then also called ``Gaussian effective potential''. Some
references are given in \cite{mass}. In \cite{gro}, on the other hand,
the whole Lagrangian was scaled with a constant.

It has been suggested in the literature \cite{dunjo} that one should
start the procedure with the most general free Lagrangian. This would
lead to a non-local quadratic action containing an arbitrary form
factor. It is also believed that the behavior in the ultraviolet
region should be identical to that of usual perturbation theory if the
form factor approaches one for large momenta. We will see that the
situation is in fact more complicated in our case.

In the present paper we apply this idea to QCD, for simplicity
without quarks for the present. The first central point, treated in
sect. 2, is the construction of a Lagrangian which is gauge invariant
for every $\delta $. This can be done by connecting the field tensors,
now taken apart to different space time points in the non local
Lagrangian, by a path ordered string. Gauge fixing can be done in
analogy to the familiar case. The action becomes an infinite series in
$\delta $ which coincides with the usual one for $\delta =1$. For
explicit calculations one expands the action up to the desired order.

In sect. 3 we give the general formula for the gluonic vacuum
polarization. The transversality can be explicitely checked, thus
confirming the manifest gauge invariance of the formalism. In sect. 4
we give the corresponding expressions for the ghost self-energy and
the ghost-gluon vertex.

Up to this point the approach is quite general and the form factor,
introduced when splitting the Lagrangian, essentially arbitrary.
In sect 5. we discuss some conceptional questions which arise when
one tries to fix the form factor in an optimal way. Due to
divergences this is a non-trivial task. In sect 6. we therefore
specialize to a scale-invariant ansatz and concentrate on a
calculation of the anomalous dimensions and the $\beta $-function.
This is much simpler because only the divergent contributions have to
be extracted.

Due to the complexity of the problem we have not yet achieved a
break-through in the present paper. We believe, however, that we have
demonstrated the feasibility and the potential of the method. With
some optimism one may hope that it can become a new alternative
approach to quantum field theories. In the conclusion we discuss, how
non-perturbative results could be obtained in future applications.

\setcounter{equation}{0}\addtocounter{saveeqn}{1}%

\section{The action }

Our starting point is the classical Yang-Mills action

\begin{equation} S^{(cl)} = -\frac{ 1}{4 }\int  F_{\mu \nu }^a (x)
F^{\mu \nu a} (x) dx.\end{equation}
The only modification at this point is that we replace the bare
coupling constant $g_0$ in $F_{\mu \nu }^a$ by $\delta \, g_0$. The
same replacement is made in the definition of gauge transformations.
This is necessary in order to obtain gauge invariance for any $\delta
$, which is of central importance. Thus we put

\begin{equation} F_{\mu \nu}^a  =  \partial _\mu A_\nu ^a- \partial
_\nu A_\mu ^a + \delta g_0 f^{abc}A_\mu ^b A_\nu ^c
\equiv G_{\mu \nu }^a + \delta g_0 C_{\mu \nu }^a. \end{equation}
For later convenience we have introduced the notations $G_{\mu \nu }^a
$ and $C_{\mu \nu }^a$ for the abelian part and the
commutator term of $F_{\mu \nu }^a$, respectively.

In order to start the $\delta $-expansion with a general non-local
free action we take the two factors $F_{\mu \nu }^a(x)$ in (2.1) apart
to different space time points $x,y$ and introduce a Lorentz invariant
form factor $K(x-y)$. To do this in a gauge invariant way one has to
connect the points $x,y$ by a path ordered octet exponential $U(x,y)$:

\begin{equation} U(x,y) = P\exp [i\delta g_0 \int_y^xT^aA_\mu
^a(z)dz^\mu ], \end{equation}
with $(T^a)_{bc} = -i f^{abc}$ the representation matrices in the
octet representation. Technically this is much simpler than using two
factors $U(x,y)$ and $U(y,x)$ in the triplet representation. So we
rewrite (2.1) as

\begin{eqnarray} S^{(cl)} & = & -\frac{ 1}{4 } \int[(1-\delta ^2)K(x-
y) + \delta ^2 \delta (x-y)] F_{\mu \nu }^a(x) U^{ab}(x,y) F^{\mu \nu
b} (y) dx dy \nonumber\\ & = & S^{(cl)}_0 + S^{(cl)}_I, \end{eqnarray}
with
\begin{equation} S^{(cl)}_0 = -\frac{ 1}{4 }\int K(x-y) G_{\mu \nu}^a
(x) G^{\mu \nu a} (y)  dx dy  . \end{equation}
Clearly (2.4) is gauge invariant for any $\delta $ and coincides with
the original action (2.1) for $\delta =1$. We used $\delta ^2$ and not
$\delta $ in the first two terms of (2.4) in order to keep the analogy
with usual perturbation theory, where the first correction to the
wave-function renormalization is also quadratic in the coupling
constant. Furthermore the Lagrangian (2.4) is symmetric with respect
to the transformation $A_\mu \rightarrow -A_\mu ,\quad \delta
\rightarrow -\delta $. The free action $S^{(cl)}_0 $ is the part of
$S^{(cl)}$ which survives for $\delta =0$, while $S^{(cl)}_I $ denotes
all the rest.

Next we perform gauge fixing and quantization with the help of the

Faddeev-Popov procedure, slightly adopted to our case. We introduce
the covariant gauge

\begin{equation} F[A^{\mu a}(x)] = \int K(x-y)[\partial ^\mu A_\mu
^a(y) - \chi ^a (y)]dy. \end{equation}
The ghost Lagrangian thus becomes

\begin{equation} S^{(ghost)} = -\int  \bar{\omega }^a(x) K(x-y)
\{\Box\omega ^a(y) - \delta g_0f^{abc}\partial ^\mu [A_\mu ^b(y)\omega
 ^c(y)]\}dxdy.  \end{equation}
At the end of the gauge fixing procedure we integrate over the
auxiliary fields $ \chi ^a $ with the weight function

\begin{equation} \exp [-\frac{ i }{2 \alpha } \int \chi ^a (x) K(x-y)
\chi ^a (y) dx dy]. \end{equation}
The result then combines with the terms involving $\partial ^\mu
\partial ^\nu A_\nu ^a$ in the usual way. We thus arrive at the
action $S=S_0 + S_I$ with

\begin{eqnarray} S_0 & = & \frac{ 1}{2 }\int A_\mu ^a(x) K(x-y) \{\Box
A^{\mu a}(y) +(1/\alpha -1)\partial ^\mu \partial ^\nu A_\nu ^a(y)\}
dx dy \nonumber\\
& & -\int  \bar{\omega }^a(x)K(x-y)\Box \omega ^a(y) dxdy,
\end{eqnarray}
\begin{equation} S_I = S^{(cl)}_I -\delta g_0 f^{abc}\int \partial
^\mu \bar{\omega }^a(x)K(x-y)A_\mu ^b(y)\omega ^c(y) dx dy.
\end{equation}
The free gluon propagator will therefore have the form

\begin{equation} D_{\mu \nu }^{ab}(q) = \delta
^{ab}D_{\mu \nu }(q) = \delta ^{ab}D(q)(g_{\mu \nu } -
\xi \frac{ q_\mu q_\nu }{q^2 }),\quad \mbox{with}\quad \xi
=1-\alpha .  \end{equation}
Here

\begin{equation} D(q) = \frac{1}{(q^2+i\epsilon )K(q)},
\end{equation}
with $K(q)$ the Fourier transform of $K(x)$. The free ghost propagator
is also given by $D(q)$. In fact the form factor in the ghost gluon
vertex will cancel against the one in the ghost propagator in all
ghost loops. This is due to the fact that our Faddeev-Popov
determinant is just the usual one, multiplied with a field-independent
term. The form (2.6) which leads to (2.7) has, however, the special
virtue that the total action given below is invariant under the usual
BRS transformation.

We mention that the special case $K=1/\zeta $ with constant $\zeta $
would lead back to the ansatz in \cite{gro} which is, however, only
useful if one wants to connect renormalized with unrenormalized
quantities. If one expresses renormalized quantities through
renormalized parameters, a constant $\zeta $ drops out.

For the explicit calculations we expand the string $U(x,y)$ in
powers of $\delta $ and insert the series into (2.4). We thus obtain
the following expansion of the action, which we will need only up to
order $\delta ^2$ here.

\begin{equation} S = S_0 + \delta S^{(1)} + \delta ^2 S^{(2)} +
\cdots .\end{equation}
$S_0$ was already given in (2.9), the interaction terms below are
classified in an obvious notation, where the indices $ins,C,S,Gh$
denote the origin from insertions (the term $-\frac{\delta ^2}{4} \int
[\delta (x-y) -K(x- y)] G_{\mu \nu }^a(x) G^{\mu \nu a}(y) dxdy$
present in (2.4)), commutator, string, or ghost interactions
respectively. One finds

\begin{eqnarray} S^{(1)}_C & = & - g_0 f^{abc}\int \partial _\mu
A_\nu ^a (x) K(x-y)A^{\mu b}(y) A^{\nu c}(y)dxdy,\nonumber\\
S^{(1)}_S & = & \frac{g_0 }{2 } f^{abc} \int \partial _\mu A_\nu
^a(x)K(x-y) (x-y)^\rho A_\rho ^b(sx+(1-s)y)[\partial ^\mu A^{\nu c}(y)
- \partial ^\nu A^{\mu c}(y)]dxdyds,\nonumber\\
S^{(1)}_{Gh} & = &  - g_0 f^{abc}\int  \partial ^\mu \bar{\omega
}^a(x)K(x-y)A_\mu ^b(y) \omega ^c(y)dxdy.\end{eqnarray}

\begin{eqnarray} S^{(2)}_{ins} & = & -\frac{ 1}{2 } \int \partial
_\mu A_\nu ^a(x) [\delta (x-y) - K(x- y)] [\partial ^\mu A^{\nu a}(y)
- \partial ^\nu A^{\mu a} (y)]dx dy,\nonumber\\
S^{(2)}_{CC} & = &-  \frac{ g_0^2}{4 } f^{abc}f^{ade}\int A_\mu
^b(x)A_\nu ^c(x)K(x-y) A^{\mu d}(y)A^{\nu e}(y)dxdy,\nonumber\\
S^{(2)}_{CS} & = &g_0^2 f^{abc}f^{ade} \int A_\mu ^b(x)A_\nu ^c(x)
K(x-y)(x-y)^\rho A_\rho ^d(sx+(1-s)y) \partial ^\mu A^{\nu e} (y)
dxdyds,\nonumber\\
S^{(2)}_{SS} & = & - \frac{ g_0^2}{2 } f^{abc}f^{ade}
\int \partial _\mu A_\nu ^b(x)
K(x-y) \Theta(s-s') (x-y)^\rho A_\rho ^c(sx+(1-s)y) \nonumber\\
& & (x-y)^\sigma A_\sigma ^d(s'x+(1-s')y) [\partial ^\mu A^{\nu
e}(y)-\partial ^\nu A^{\mu e}(y)]dxdydsds'. \end{eqnarray}

Due to the strings, the action is in fact an infinite series in
$\delta $. This is a necessary consequence of the non-locality and the
gauge invariance of the approach. In any finite order we will, of
course, only need a finite number of terms.

The vertices of the action can be visualized by slightly modified
Feynman graphs which are shown in fig. 1 in momentum space. The
propagators refer to $D(q)$ now, the presence of a thick line denotes
a factor $K(q)$. If such a factor appears in an internal line, it
results in a total propagator $D(q) K(q)=1/(q^2+i\epsilon )$, i.e. the
usual free propagator. Thick lines with $n$ gluon lines attached at
the interior of the line denote the $n$-th order expansion of the
string $U(x,y)$. The graphs make the structure of the action more
transparent, we did, however, not set up general modified Feynman
rules here, but preferred the direct one-loop calculation.

\newpage

\setcounter{equation}{0}\addtocounter{saveeqn}{1}%

\section{Vacuum polarization }

We start with the one particle irreducible contributions to
the full gluon propagator which make up the vacuum polarization
$\Pi _{\mu \nu}$. Due to the nonlocality of $S_0$, it is most
convenient to write down the path integral representation for the
propagator, expand it with respect to $\delta $ , perform the Gaussian
integrations and calculate the contractions in momentum space. The
whole calculation can now be done using familiar methods. The result,
written in $d$ dimensions and for $N_c$ colors, has the form

\begin{equation} \Pi_{\mu \nu }(q) = \delta ^2[K(q)-1] (q^2g_{\mu \nu
}- q_\mu q_\nu ) -\frac{ i \delta ^2 g_0^2N_c} {(2\pi )^d} \int \pi
_{\mu \nu }(q,k) d^dk. \end{equation}
The first contribution at the rhs of (3.1) stems from the insertion
$S^{(2)}_{ins}$ in (2.15). The integrand $\pi_{\mu \nu }(q,k)$ is a
sum of 20 terms which we denote by $\pi_{\mu \nu }^{(j)}$. Terms (1)
to (14) are loop terms arising from $S^{(1)} S^{(1)}$ in (2.14):
(1)-(4) from $S^{(1)}_C S^{(1)}_C$, (5)-(9) from $S^{(1)}_C
S^{(1)}_S$, (10)-(13) from $S^{(1)}_S S^{(1)}_S$, and (14) from
$S^{(1)}_{Gh} S^{(1)}_{Gh}$.

Terms (15)-(20) originate from the
second-order part $S^{(2)}$ in (2.15): Term (15) is the tadpole from
the commutator term $S^{(2)}_{CC}$, (16),(17) stem from $S^{(2)}_{CS}$
, and (18)-(20) from $S^{(2)}_{SS}$ . Wherever possible, we simplified
the expressions by using relation (2.12). The contributions are shown
in graphical representation in fig. 2. The explicit forms read:

\begin{eqnarray}
\pi _{\mu \nu }^{(1)} & = & \{ q^2g_{\mu \nu } - q_\mu q_\nu
-\frac{\xi } {k^2}[qk(qkg_{\mu \nu }-q_\mu k_\nu )+ (q^2 k_\mu
-qkq_\mu )k_\nu )] \nonumber\\
& & +\frac{ \xi ^2q(q-k)}{k^2(q-k)^2 }k_\mu [q^2k_\nu -qkq_\nu ]
\}  D(k)D(q-k)K^2(q) ,\nonumber\\
\pi _{\mu \nu }^{(2)} & = & 2\{ qkg_{\mu \nu } - q_\mu k_\nu -\xi
\frac{ k(q-k)}{(q-k)^2 }[q(q-k)g_{\mu \nu }-q_\mu (q-k)_\nu ]\}
D(q-k)\frac{K(q)}{k^2 }, \nonumber\\
\pi _{\mu \nu }^{(3)} & = & -(d-1)k_\mu (q-k)_\nu \frac{ 1}{k^2(q-k)^2
}, \nonumber\\
\pi _{\mu \nu }^{(4)} & = & \{ k^2 g_{\mu \nu } +(d-2) k_\mu k_\nu -
\frac{\xi }{(q-k)^2 }[(k(q-k))^2 g_{\mu \nu } -2k(q-k)q_\mu k_\nu
\nonumber\\
& & +(q^2-k^2)k_\mu k_\nu ]\} D(q-k)\frac{K(k)}{k^2},\nonumber\\
\pi _{\mu \nu }^{(5)} & = & \{q^2k_\mu -qkq_\mu \}
D(k)D(q-k) K(q)\int K_\nu
(k-sq)ds,\nonumber\\
\pi _{\mu \nu }^{(6)} & = & 2\{[q(q-k)g_{\mu \nu }-q_\mu (q-k)_\nu ]
(q_\rho -\xi \frac{ qk}{k^2 }k_\rho )-(q^2k_\mu -qkq_\mu )(g_{\nu \rho
}-\xi \frac{ k_\nu k_\rho }{k^2})\}\nonumber\\ & &
D(k)D(q-k)K(q)\int K^\rho (q-
sk)ds,\nonumber\\
\pi _{\mu \nu }^{(7)} & = & -2\{(d-2)k(q-k)k_\mu +k^2(q-k)_\mu
\}\frac{ D(q-k)}{k^2}\int K_\nu (k-sq)ds,\nonumber\\
\pi _{\mu \nu }^{(8)} & = & 2\{ [q(q-k)g_{\mu \nu }-q_\mu (q-k)_\nu ]
[(q-k)_\rho - \xi \frac{k(q-k)}{k^2 }k_\rho ] \nonumber\\
& & -(q-k)_\mu [q(q-k)g_{\nu \rho }-(q-k)_\nu q_\rho ]\nonumber\\
& & -\frac{ \xi }{k^2 }(q-k)_\mu [qkq_\nu -q^2k_\nu ]k_\rho \}
 \frac{ D(k)} {(q-k)^2 }\int K^\rho (q -sk)ds, \nonumber\\
\pi _{\mu \nu }^{(9)} & = & 2(q-k)_\mu \{kqg_{ \nu \rho }- k_\nu
q_\rho \} \frac{ D(q-k)}{k^2 }\int K^\rho (q-sk)ds, \nonumber\\
\pi _{\mu \nu }^{(10)} & = &
\frac{ 1}{2 }\{(d-2)(k(q-k))^2+k^2(q-k)^2\}D(k) D(q-
k) \nonumber\\
& & \int\int K_\mu (sq-k) K_\nu
(s'q-k) dsds',\nonumber\\
\pi _{\mu \nu }^{(11)} & = & -2 \{ k(q-k)[(q-k)qg_{\mu \rho } -
(q-k)_\mu q_\rho ] + [qkq_\mu -q^2k_\mu] (q-k)_\rho \} \nonumber\\
& & D(k)D(q-k)\int\int
K^\rho (q-sk) K_\nu (k-s'q)
dsds',\nonumber\\
\pi _{\mu \nu }^{(12)} & = & -[kqg_{\mu \rho }-k_\mu q_\rho ] \{(q-
k)qg_{\nu \sigma  }- (q-k)_\nu q_\sigma \} D(k)D(q-k)
\nonumber\\ & & \int\int K^\rho (q-sk)K^\sigma
((1-s')q+s'k)dsds',\nonumber\\
\pi _{\mu \nu}^{(13)} & = & (g_{\rho \sigma }-\xi \frac{ k_\rho
k_\sigma }{k^2 })\{ q(q-k)[q(q-k)g_{\mu \nu } - q_\mu (q-k)_\nu ]
\nonumber\\
& & +[q^2(q-k)_\mu -q(q-k)q_\mu ] (q-k)_\nu )\}\nonumber\\ & &
D(k)D(q-k)\int\int K^\rho (q-sk)
K^\sigma (q-s'k) dsds',\nonumber\\
\pi _{\mu \nu }^{(14)} & = & \frac{ k_\mu (q-k)_\nu
}{k^2(q-k)^2 }, \nonumber\\
\pi _{\mu \nu }^{(15)} & = & -\{ (d-1-\xi
)g_{\mu \nu} +\xi \frac{ (q- k)_\mu (q-k)_\nu }{(q-k)^2 }\}
D(q-k)K(k),\nonumber\\
\pi _{\mu \nu }^{(16)} & = & -2\{g_{\mu \nu }q_\rho -g_{\mu \rho}
q_\nu - \frac{ \xi } {k^2} (qkg_{\mu \nu }-q_\mu k_\nu )k_\rho \}
D(k) \int K^\rho (q+sk)ds,\nonumber\\
\pi _{\mu \nu }^{(17)} & = &
-2(d-1)k_\mu D(k)\int K_\nu (k+sq)ds,\nonumber\\
\pi _{\mu \nu}^{(18)} & = & -[q^2g_{\mu \nu }-q_\mu q_\nu
](g_{\rho \sigma }-\xi \frac{ k_\rho k_\sigma } {k^2 })
D(k)\int \int \Theta(s-s')K^{\rho \sigma }(q+(s-
s')k)dsds',\nonumber\\
\pi _{\mu \nu }^{(19)} & = & -(d-1) k^2D(k) \int \int
\Theta(s-s')K_{\mu \nu }(k+(s-s')q)dsds', \nonumber\\
\pi _{\mu \nu}^{(20)} & = & 2[kq g^\rho _\mu -k_\mu q^ \rho ]D(k) \int
\int \Theta(s-s')K_{\rho \nu }((1-s')q+sk)dsds'.\end{eqnarray}
The integrations over $s$ or $s'$ in the string terms, run from 0
to 1. We introduced the short hand notations $K^\mu (k) \equiv
\partial ^\mu K(k)$ and $K^{\mu \nu }(k) \equiv \partial
^\mu \partial ^\nu K(k)$. The string terms contain only
derivatives of $K$ because they have to vanish for the usual
local form $K= const$.

Our procedure in the last section guarantees that $\Pi _{\mu \nu}$ is
transversal. We check this by contracting $\Pi_{\mu \nu }$ with $q^\mu
q^\nu $. We denote the corresponding integrands by $\pi
_l^{(1)},\cdots,\pi _l^{(20)}$. The non vanishing terms are

\begin{eqnarray}
\pi _l^{(3)} & = & -(d-1)qkq(q-k)\frac{
1}{k^2(q-k)^2 }, \nonumber\\
\pi _l^{(4)} & = & \{ q^2 k^2+ (d-2)(qk)^2 -\xi \frac{
k^2[q^2k^2-(qk)^2]}{(q-k)^2 } \}\frac{D(q- k)K(k)}
{k^2},\nonumber\\
\pi _l^{(7)} & = &
-2\{(d-2)qk\;k(q-k)+k^2q(q-k)\}\frac{D(q-k)}{k^2}\int
q^\nu K_\nu  (k-sq) ds\nonumber\\
& = & -2\{(d-2)qk\;k(q-k)+k^2q(q-k)\}[\frac{D(q- k)K
(k)}{k^2} -\frac{ 1}{k^2(q-k)^2 }],\nonumber\\
\pi _l^{(10)} & = & \frac{ 1}{2 }\{(d-2)(k(q-k))^2+k^2(q-k)^2\}
D(k) D(q-k) \nonumber\\ & & \int \int q^\mu
K_\mu (sq-k) q^\nu K_\nu (s'q-k)ds ds'\nonumber\\ & = &
\{(d-2)(k(q-k))^2+k^2(q-k)^2\}[\frac{D(q- k) K(k)}
{k^2} -\frac{ 1}{k^2(q-k)^2 }]\nonumber\\
\pi _l^{(14)} & = & \frac{qkq(q-k)}{k^2 (q-k)^2}, \nonumber\\
\pi _l^{(15)} & = & -\{ (d-1-\xi )q^2 + \xi \frac{ (q(q-k))^2}{(q-k)^2
}\} D(q-k)K(k),\nonumber\\
\pi _l^{(17)} & = & -2(d-1)qkD(k)\int q^\nu K_\nu
(k+sq)ds,\nonumber\\
& = & 2(d-1)k^2q(q-k)[\frac{D(q- k)K(k)} {k^2} -\frac{
1}{k^2(q-k)^2 }],\nonumber\\
\pi _l^{(19)} & = & -(d-1)k^2D(k) \int \int
\Theta(s-s')q^\mu q^\nu K_{\mu \nu } (k+(s-s')q)dsds'\nonumber\\
& = & -(d-1)k^2(q-k)^2[\frac{D(q- k)K(k)} {k^2}
-\frac{ 1}{k^2(q-k)^2 }].\end{eqnarray}
The second forms for the string terms $\pi _l^{(7)},\pi _l^{(10)},\pi
_l^{(17)},\pi _l^{(19)}$ which do no longer contain integrations over
$s,s'$ are obtained as follows: E.g. in $\pi _l^{(19)}$ use $q^\mu
q^\nu K_{\mu \nu }(k+(s-s') q) = -(d/ds) (d/ds')K(k+(s-s')q)$, perform
the integrations over $s$ and $s'$, and substitute $k\rightarrow k-p$
where appropriate. The cancellation of the sum $\sum _j\pi _l^{(j)}$
now happens in the following way: All the $\pi _l^{(j)}$ contain
either the factors $D(q- k)K(k)/k^2$ or the products of the free
propagators $1/k^2(q- k)^2$. The terms of the first type add up to
zero, the sum of the latter ones may be written as

\begin{equation} -(d-2)\{\frac{ q(q-k)}{(q-k)^2 }+\frac{ qk}{k^2 }\}.
\end{equation}
Obviously this expression vanishes after integration over $k$.

Due to the transversality of $\Pi_{\mu \nu }$ we may write

\begin{equation} \Pi _{\mu \nu }(q) = (q^2 g_{\mu \nu } -  q_\mu
q_\nu )\Pi(q^2) .\end{equation}
$\Pi (q^2)$ has the form

\begin{equation} \Pi (q^2) = \delta ^2[K(q)-1] -\frac{ i \delta ^2
g_0^2N_c} {(2\pi )^d} \int \pi (q,k) d^dk. \end{equation}
It is easy to obtain the integrands $\pi ^{(j)}(q,k)$
appearing in $\Pi(q^2)$. Since we have already checked the
transversality for the sum in (3.2), one can simply
calculate the trace: $\pi =\sum _j\pi ^{(j)} =\sum _j\pi_\mu
^{(j)\mu }/((d-1)q^2) $.

Proceeding further in the usual way one gets for the gluon propagator

\begin{equation}
\delta ^{ab} \Delta_{\mu \nu }(q) = \delta ^{ab}\{ D_{\mu \nu }(q)
+ D_{\mu \rho }(q)\Pi ^{\rho \lambda}(q) D_{\lambda \nu }(q) +
D\Pi D\Pi D +\cdots\}. \end{equation}
In matrix notation the geometrical series becomes

\begin{equation} (\Delta ) = [(D)^{-1} - (\Pi)]^{-1}.
  \end{equation}
This gives

\begin{equation} \Delta_{\mu \nu }(q) = \Delta(q)
 [g_{\mu \nu }-\tilde{\xi }(q^2)\frac{q_\mu q_\nu}{q^2}]
,\end{equation}
with

\begin{equation} \Delta(q) = \frac{1} {(q^2+i\epsilon ) [K(q)- \Pi
(q)]} \end{equation}
and a modified $q^2$-dependent $\tilde{\xi }(q^2)$ which is of no
interest here. The gluon wave-function renormalization constant
$Z_{gluon}$ , defined at the renormalization scale $-Q^2>0$, is
obtained from $ \Delta(-Q^2) = Z_{gluon}/(-Q^2)$. In order
$\delta ^2$ this gives

\begin{equation} Z_{gluon} = [1+\Pi (-Q^2)/K(-Q^2)]/K(-Q^2).
\end{equation}

\setcounter{equation}{0}\addtocounter{saveeqn}{1}%

\section{ Ghost self-energy and vertex function}

A calculation of the renormalized coupling constant from the gluon
three-point function would be rather complicated. For this one would
need the expansion of the action up to order $\delta ^3$. Furthermore
many mixing terms between $S^{(1)}$ and $S^{(2)}$ would show up.
Therefore we will use the ghost-gluon vertex instead, which is much
simpler.

We start with the ghost self-energy $\Sigma(q^2)$.
It consists only of the graph (21) (fig. 3) and reads

\begin{equation} \Sigma (q)/q^2 = -\frac{ i\delta ^2g_0^2N_c}
{(2\pi )^d } \int\sigma ^{(21)}(q,k) d^dk, \end{equation}
with
\begin{equation} \sigma ^{(21)}(q,k) = -[q(q-k) - \xi \frac{ qk(q-
k)k}{k^2 }]\frac{K(q)D(k)} {q^2(q-k)^2}. \end{equation}
It is connected to the ghost propagator $\Delta _{ghost}(q)$ by

\begin{equation} \Delta _{ghost}(q) = \frac{1}{(q^2+i\epsilon )[K(q) -
\Sigma(q)/q^2 ]}. \end{equation}
Defining the ghost wave-function renormalization constant $Z_{ghost}$
by $ {\Delta }_{ghost}(-Q^2) = Z_{ghost}/(-Q^2)$
one has in order $\delta ^2$

\begin{equation} Z_{ghost} = [1 + \Sigma(-Q^2)/(-Q^2)K(-Q
^2)]/K(-Q^2). \end{equation}
We next consider the ghost-gluon vertex function $\Gamma _\mu
^{abc}(p,q)$, where $p$ and $q$ denote the momenta of the incoming and
outgoing ghost line with color indices a and b. We write it in the
form

\begin{equation} \Gamma _\mu ^{abc}(p,q) = f^{abc}\Gamma _\mu (p,q) =
f^{abc}[K(q)q_\mu -\frac{i\delta ^2g_0^2N_c} {(2\pi )^d} \int \gamma
_\mu (p,q,k)d^dk]. \end{equation}
There are seven contributions (fig. 4). Graphs (22) - (25) are
generalizations of the usual vertex graphs, (26) - (28) are string
terms. For general momenta the expressions are rather lengthy. We
therefore specialize to the case $p=q$, i.e. vanishing gluon four
momentum. This is sufficient for our purpose. The vertex
function simplifies to $\Gamma_\mu (q,q) =q_\mu \Gamma (q)$ with

\begin{equation} \Gamma (q) = K(q) -\frac{i\delta ^2g_0^2N_c} {(2\pi
)^d} \int \gamma (q,k)d^dk. \end{equation}
Here $\gamma (q,k) = q^\mu \gamma _\mu (q,q,k)/q^2$ is a sum of seven
terms $\gamma ^{(j)}(q,k)$, only four are different from zero. They
read:

\begin{eqnarray}
\gamma ^{(22)} & = & [q(q-k) - \xi \frac{ qk k(q-
k)}{k^2}]\frac{q(q-k)D(k)K(q)} {2q^2(q-k)^4 },\nonumber\\
\gamma  ^{(24)} & = & \xi [q^2k^2-(qk)^2]\frac{q(q-k)D(q-k)K(q)}
{2q^2k^2(q-k)^4},\nonumber\\
\gamma ^{(25)} & = & [q^2k^2-(qk)^2]\left\{1+\xi \frac{ k(q-k)}{ (q-
k)^2 }\right\}\frac{D(q-k)K(q)} {2q^2k^2(q-k)^2 } ,\nonumber\\
\gamma ^{(28)} & = & -[q^2k^2-(qk)^2] \frac{D^2(q-k) K(q) q_\mu K^\mu
(q-k) } {2q^2k^2 },\nonumber\\
\gamma  ^{(23)} & = & \gamma  ^{(26)} = \gamma  ^{(27)} =0.
\end{eqnarray}
The vertex renormalization constant $\tilde{Z}_{vertex}$
is directly related to (4.6):

\begin{equation} \tilde{Z}_{vertex}^{-1}=\Gamma (-Q^2).\end{equation}

\setcounter{equation}{0}\addtocounter{saveeqn}{1}%

\section{Conceptual questions }

The derivation of the formulae in the last section, though somewhat
tedious, was essentially straightforward. The real problem starts,
when the expressions are to be evaluated and, in particular, when one
has to find a principle for the ``optimal'' choice of the form factor
$K(k)$. We will discuss some of the problematics here before entering
more detailed calculations.

The evaluation can be performed in the following way. We
assume that both $K(k)$ and $D(k)=1/(k^2+i\epsilon )K(k)$ satisfy
spectral representations with roughly the same behavior for large
$k^2$ as in the free case where $K(k)=1,D(k) = 1/(k^2+i\epsilon )$.
Therefore we write down a once subtracted K\"allen-Lehmann
representation for $K$, with the subtraction point chosen at infinity
for convenience. For $D$ we assume an unsubtracted dispersion
relation. Thus put

\begin{equation} K(k^2) = 1 + \int \frac{ \bar{\kappa } (\mu ^2) d\mu
^2}{k^2-\mu ^2+i\epsilon },\quad D(k^2)=\int \frac{ \bar{\rho }(\mu
^2)d\mu ^2}{k^2-\mu ^2+i\epsilon  }. \end{equation}
Due to (2.12) the spectral functions $\bar{\kappa }$ and $\bar{\rho }$
are not independent. One could scale $K(k)$ with a constant $C$ and $
D(k)$ with $1/C$, respectively. The normalization constant $C$
cancels, however, in all the loop graphs contributing to the gluon
vacuum polarization, the ghost self-energy, and the vertex function.
It only enters in the insertion to the vacuum polarization (first term
on the rhs in (3.1)). In the following the normalization constant $C$
will be of no importance, therefore we choose it equal to 1.

The momentum integrations $\int \cdots d^dk$ in the expressions of the
previous sections can now be performed in the usual way, although this
becomes somewhat ugly for some of the string terms. At the end one is
left with some integrations over the spectral functions $\bar{\kappa
}$ and $\bar{\rho }$. After having calculated, say the vacuum
polarization, the ghost self-energy, and the ghost-gluon vertex in
this way, one has to decide how to choose the ``optimal'' spectral
function $\bar{\kappa }$, i.e. the input function $K(x- y)$ in our
ansatz (2.4). Two well known and successful principles suggest
themselves: The principle of fastest apparent convergence (FAC)
postulates that the considered quantity $Q$ does not change when going
to a higher order of perturbation theory, i.e. $Q_n \stackrel{!}{=}
Q_{n-1}$. The principle of minimal sensitivity (PMS) requires that the
quantity be stationary with respect to an arbitrary parameter $\lambda
$, i.e. $\partial Q_n /\partial \lambda \stackrel{!}{=}0$. In our case
PMS would not simply lead to an extremal problem but to a variational
problem, because, instead of a single parameter $\lambda $, we have an
arbitrary function $\bar{\kappa }$ at our disposal.

In a finite theory one could apply FAC or PMS to, say, the propagator
and thus obtain a non-perturbative solution. In fact this can be done
very easily in a toy model like four-dimensional $\Phi ^3$ theory (I
thank I. Solovtsov for suggesting this simple exercise and N.
Brambilla and A. Vairo for an enlightening discussion of the result).
FAC leads to an integral equation of the Dyson-Schwinger type which
can easily be solved by iteration. PMS, on the other hand, is not
applicable in one-loop order, because the propagator becomes
independent of the spectral function.

Fundamental conceptional questions arise, however, if one tries to
apply these ideas to field theories with divergences as in the
present case. It appears natural to apply the above principles (FAC or
PMS, respectively) to renormalized quantities, expressed by
renormalized parameters. But the situation is more subtle here than in
usual perturbation theory. The reason is, that renormalized quantities
are not necessarily finite for a general function $K$! This is rather
obvious, e.g. from the expressions in (3.2): The divergent terms have
different $q^2$-dependent factors in front and thus will not cancel in
the differences appearing, say, in renormalized propagators.
Technically, the reason is that our action (2.9), (2.10), though gauge
invariant for every $\delta $, is not renormalizable if $\delta $ is
considered as the coupling constant.

One way to overcome the problem of divergences would be to simply
drop the divergent contributions in renormalized quantities. From
ordinary perturbation theory we know that they have to be absent, so
also in our case they have to cancel when the whole perturbation
series in $\delta $ is summed up. The removal of the divergent terms
is certainly not unique, analogous to the ambiguities of
renormalization schemes. In finite orders the results will depend on
the detailed prescription, the exact result should, however, be
independent of it. The remaining finite quantities could then be
determined by using FAC or PMS. We shall not investigate this
possibility here but proceed with a calculation of the anomalous
dimensions and the $\beta $-function, where the problem of removing
divergences does not show up.

\setcounter{equation}{0}\addtocounter{saveeqn}{1}%

\section{Anomalous dimensions and $\beta $-function }

The $\beta $-function describes the scaling behavior of the quantized
field theory which differs from the naive expectations from classical
scale invariance. In order to maintain the classical scale invariance
in every order of the $\delta $-expansion we make a scale-invariant
ansatz for the spectral functions $\bar{\kappa }$ (dimension 0) and
$\bar{\rho }$ (dimension -2). The only scale available is the external
momentum $q$ in the propagators or the vertex function, respectively.
With $Q^2 = -q^2>0$ the euclidean squared momentum, we therefore put

\begin{equation} \bar{\kappa }(\mu ^2) = \kappa (\mu ^2/Q^2) \equiv
\kappa (m^2). \end{equation}
The function $\kappa $ as well as the integration variable $m^2 = \mu
^2/Q^2$ are now dimensionless. From (5.1) we get

\begin{equation} K(k^2) = 1 + \int \frac{\kappa (m^2)}
{k^2/Q^2-m^2+i\epsilon }dm^2. \end{equation}
In particular we have
\begin{equation} K(-Q^2) = 1 - \int \frac{\kappa (m^2)} {1+m^2}dm^2,
\end{equation}
i.e. $K(-Q^2)$ becomes independent of $Q^2$. Therefore the
corresponding factors present e.g. in $\pi ^{(1)},\pi ^{(2)},\cdots $
are not differentiated when calculating the $\beta $-function and the
divergent contributions disappear as usual.

From simple dimensional analysis in $d = 4 -2\epsilon $ dimensions the
integrals appearing in (3.6),(4.1),(4.6) are proportional to
$(Q^2)^{-\epsilon }= 1-\epsilon \ln\,Q^2$. Only the divergent terms
$\sim 1/\epsilon $ survive after applying the operator $Q^2 d/dQ^2$.
By naive power counting in $k$ one would conclude that all the string
contributions in the vacuum polarization (3.2), with the exception of
$\pi ^{(7)},\pi ^{(17)},\pi ^{(19)}$ should be convergent. This is,
however not true. The string parameter $s$ has to be integrated from 0
to 1, and for $s=0$ most of the terms are divergent again by power
counting. In fact, a careful investigation allows to extract the
divergent contribution arising from the region of small $s$. In the
appendix we give some technical details, how this can be done in a
rather straightforward way.

The propagators $D$ will be treated by expressing $D$ through $K$ by
using (2.12), introducing (5.1) (or (6.2)) for $K$ and expanding with
respect to the integral:

\begin{equation} D(k^2) = [(k^2 +i\epsilon )K(k^2)]^{-1} = \frac{ 1}
{(k^2 +i\epsilon )} [1 - \int \frac{ \bar{\kappa }(\mu ^2)d\mu ^2}
{k^2-\mu ^2+i\epsilon }] + O(1/k^6). \end{equation}
The terms of order $1/k^6$ lead to finite integrals and are therefore
irrelevant.

We write the renormalization constants, derived in (3.11), (4.4),
(4.8) in the form

\begin{eqnarray} Z_{gluon} & = & \frac{1}{K} [1 + \delta ^2(K-1)/K +
\frac{ \delta ^2g_0^2N_c}{(4\pi )^2K }(\frac{ 1}{\epsilon }-\ln Q^2)
\hat{\pi }],\\
Z_{ghost} & = & \frac{1}{K} [1 + \frac{ \delta ^2g_0^2N_c}{(4\pi )^2K}
(\frac{ 1}{\epsilon }-\ln Q^2) \hat{\sigma }],\\
\tilde{Z}_{vertex}^{-1}  & = & K [1 + \frac{ \delta
^2g_0^2N_c}{(4\pi )^2K }(\frac{ 1}{\epsilon }-\ln Q^2) \hat{\gamma }].
\end{eqnarray}

In the above equations we have abbreviated

\begin{equation} K\equiv K(-Q^2).  \end{equation}
We have only written down the divergent terms $\sim (1/\epsilon - \ln
Q^2)$ of the integrals $\pi ,\sigma ,\gamma $, the factors in front
have been denoted by $\hat{\pi },\hat {\sigma },\hat{\gamma }$.
The renormalized coupling constant becomes

\begin{eqnarray} g = \delta
g_0Z_{gluon}^{1/2}Z_{ghost}\tilde{Z}_{vertex}^{-1} = \frac{\delta
g_0}{\sqrt{K}}[1 + \frac{\delta ^2}{2K}(K-1) + \frac{ \delta
^2g_0^2N_c}{(4\pi )^2K }(\frac{ 1}{\epsilon }-\ln Q^2) \hat{\beta }],
\end{eqnarray}
with $\hat{\beta }=\hat{\pi }/2+\hat{\sigma }+\hat{\gamma }$. In
lowest order one has

\begin{equation} g = \delta
g_0/\sqrt{K}. \end{equation}
The various contributions to $\hat{\pi },\hat{\sigma },\hat{\gamma
},\hat{\beta }$ all have the form

\begin{equation} h + \int l(m^2) \kappa (m^2) dm^2 + \int \int
K(m^2,m'^2) \kappa (m^2)\kappa (m'^2) dm^2dm'^2,\end{equation}
i.e. they are at most quadratic in the spectral function $\kappa
(m^2)$, higher powers only contribute finite terms. Below we give the
non-vanishing contributions, where the indices correspond to the
numbers in (3.2), (4.1), and (4.7) as well as to the graphs in the
figures. All calculations can be easily done analytically, with the
exception of the quadratic contributions of the string-string graphs
(12), (13). We leave them unspecified here.

It is important to note, however, that graph (13) also has a
$\xi $-dependent term which can be calculated analytically. This will
be important for the cancellation of the $\xi $-dependence in the
$\beta $-function. The non-vanishing contributions read:

\begin{eqnarray}  h^{(1)} & = & 1-\xi /2, \nonumber\\
h^{(2)} & = & 1+\xi /2, \nonumber\\
h^{(3)} & = & -1/2, \nonumber\\
h^{(4)} & = & \xi /2, \nonumber\\
h^{(14)} & = & 1/6, \nonumber\\
h^{(21)} & = & (1+\xi /2)/2, \nonumber\\
h^{(22)} & = & (1-\xi )/8, \nonumber\\
h^{(25)} & = & 3(1-\xi )/8, \end{eqnarray}

\begin{eqnarray}
l^{(1)} & = & -\frac{  2(1-\xi /2)}{1+m^2 }, \nonumber\\
l^{(2)} & = & -\frac{  (1+\xi /2)}{1+m^2 }, \nonumber\\
l^{(6)} & = & - \frac{  2m^2+\xi }{1+m^2 } +2m^2\ln\frac{ 1+m^2}{m^2 }
, \nonumber\\
l^{(7)} & = & 4, \nonumber\\
l^{(8)} & = & -\frac{ 2m^2+\xi }{1+m^2 }+2m^2\ln\frac{ 1+m^2}{m^2 }
, \nonumber\\
l^{(9)} & = & \frac{1+ 2m^2}{1+m^2 }-2m^2\ln\frac{ 1+m^2}{m^2 } ,
\nonumber\\
l^{(17)} & = & -4, \nonumber\\
l^{(18)} & = & \frac{  2+\xi }{1+m^2 }, \nonumber\\
l^{(21)} & = & -\frac{  1+\xi /2}{2(1+m^2) }, \nonumber\\
l^{(22)} & = & -\frac{  1-\xi }{8(1+m^2) }, \nonumber\\
l^{(25)} & = & -\frac{  3(1-\xi )}{8(1+m^2) },
 \end{eqnarray}

\begin{eqnarray} K^{(1)} & = & \frac{ 1-\xi /2}{(1+m^2)(1+m'^2) },
\nonumber\\
K^{(6)} & = & (\frac{ 2m^2+\xi }{1+m^2 } -2m^2\ln\frac{ 1+m^2}{m^2})
\frac{  1}{2(1+m'^2) }+(m^2\leftrightarrow m'^2), \nonumber\\
K^{(12)} & = & \hat{K}^{(12)}(m^2,m'^2), \nonumber\\
K^{(13)} & = & \hat{K}^{(13)}(m^2,m'^2) -\frac{ \xi }{2(1+m^2)(1+m'^2)
 }. \end{eqnarray}
Summing up contributions (1) - (20) for $\hat{\pi }$, (21) for
$\hat{\sigma }$, and (22) - (25) for $\hat{\gamma }$, we obtain

\begin{eqnarray} h_{\pi } & = & 5/3 + \xi /2,
\quad l_{\pi } = -\frac{ 2m^2+\xi /2}{1+m^2 } + 2 m^2\ln\frac{
1+m^2}{m^2 } \nonumber \\ K_{\pi } & = &
\frac{1+m^2+m'^2}{(1+m^2)(1+m'^2)} -\frac{ m^2}{1+m'^2 }\ln
\frac{1+m^2}{m^2 } -\frac{ m'^2}{1+m^2 } \ln \frac{ 1+m'^2}{m'^2
}\nonumber\\ & & +\hat{K}^{(12)}(m^2,m'^2) + \hat{K}^{(13)}(m^2,m'^2)
, \nonumber \\ \end{eqnarray}
\begin{equation} h_{\sigma } = (1+\xi /2)/2,\quad l_{\sigma }= -
\frac{ 1+\xi /2}{2(1+m^2) },\quad K_{\sigma } = 0, \end{equation}
\begin{equation} h_{\gamma } = (1-\xi )/2,\quad l_{\gamma } = -\frac{
1-\xi }{2(1+m^2) },\quad K_{\gamma } = 0. \end{equation}
Furthermore we have for $\hat{\beta }=\hat{\pi }/2+\hat{\sigma }
+\hat{\gamma }$:

\begin{equation} h_\beta =11/6,\quad l_\beta = m^2\ln \frac{
1+m^2}{m^2 } -1,\quad K_\beta =\frac{ 1}{2 } K_\pi .\end{equation}

The anomalous dimensions and the $\beta $-function become

\begin{eqnarray} \gamma _{gluon} & = & \frac{ 1}{Z_{gluon} }
Q^2\frac{dZ_{gluon }}{dQ^2 } \nonumber\\ & = & -\frac{ g^2N_c}{(4\pi
)^2 }[\frac{ 5}{3 } +\frac{ \xi }{2 }+\int l_\pi (m^2)\kappa
(m^2)dm^2 \nonumber\\ & & \quad \quad \quad
+\int \int K_\pi (m^2,m'^2)\kappa (m^2)\kappa
(m'^2)dm^2dm'^2],\end{eqnarray}
\begin{equation} \gamma _{ghost} = \frac{ 1}{Z_{ghost} }
Q^2\frac{dZ_{ghost }}{dQ^2 } =-\frac{ g^2N_c}
{(4\pi )^2 }[\frac{1}{2}+\frac{\xi }{4}+\int l_\sigma (m^2)\kappa
(m^2)dm^2],  \end{equation}
\begin{eqnarray} \beta =2 Q^2\frac{ dg}{dQ^2 } & = & -\frac{ 2
g^3N_c}{(4\pi )^2} [\frac{11}{6} +\int l_\beta (m^2)\kappa
(m^2)dm^2 \nonumber\\ & & \quad \quad \quad +\int \int K_\beta
(m^2,m'^2) \kappa (m^2)\kappa (m'^2)dm^2dm'^2].\end{eqnarray}
We used (6.10) to replace $\delta g_0/\sqrt{K}$ by the renormalized
coupling constant $g$. For $\kappa (m^2) \equiv 0$ we recover the well
known results of ordinary perturbation theory in one-loop order. Note
further that the $\beta $-function is independent of the gauge
parameter $\xi $ as it should. This is a further test of the manifest
gauge invariance of the approach.

Let us next discuss whether the principle of fastest apparent
convergence (FAC) or the principle of minimal sensitivity (PMS) can be
applied, say, to $\beta $ in (6.21). An application of FAC is clearly
impossible because the present one-loop calculation is the lowest non-
trivial contribution; there is no lower order with which one could
compare. An attempt to apply PMS also fails. The variation with
respect to $\kappa (m^2)$ gives no solution at all in (6.20) which is
linear in $\kappa $. In (6.19),(6.21), on the other hand, the presence
of the linear term $\int l(m^2)\kappa (m^2) dm^2$ leads to an extremum
which is not situated at $\kappa =0$. Therefore one would not
reproduce the results of ordinary perturbation theory for small $g$.

This is not yet a problem. We know from simple toy models that the
lowest order usually gives no extremum. One needs the lowest and, at
least, the next order to get a balance between these contributions and
to find a relevant extremum. In both cases, FAC or PMS, one should
therefore go to order $\delta ^4$ for the anomalous dimensions and to
order $\delta ^5$ for the $\beta $-function. A calculation of the two
loop contributions $\sim \delta ^5g^5$ to the $\beta $-function is not
feasible at present, but one can easily look for the effect of the
contributions $\sim\delta ^5g^3$ which, of course, are the most
important ones for small coupling. These contributions arise from two
sources: First we have to consider the insertion $\delta ^2(K-1)/K$ in
$Z_{gluon}$, when expressing $\delta g_0$ by $g$. This leads to

\begin{equation} g=\frac{ \delta g_0}{\sqrt{K}}[1 + \frac{\delta
^2}{2} (1-1/K) + O(\delta ^2g_0^2)]. \end{equation}

Inverting this equation, one obtains for an arbitrary power $n$

\begin{equation} \left(\frac{\delta g_0}{\sqrt{K}}\right)^n = g^n
[1+ \frac{n\delta ^2}{2}(1/K-1) +\cdots].\end{equation}
Thus, for $\delta =1$, instead of (6.10), one should now replace

\begin{equation} \frac{\delta ^2g_0^2}{K}\Rightarrow
\frac{g^2}{K},\quad \frac{\delta ^3g_0^3}{K^{3/2}}\Rightarrow
g^3(\frac{3}{2K}-\frac{1}{2}). \end{equation}
Different replacement rules for different powers look a bit strange
but appear as a direct consequence of the expansion in $\delta $.

The second effect is that we now have to consider insertions into the
internal gluon lines in the graphs (1)-(28). Such an insertion
replaces the original propagator $D_{\mu \nu }(k) = D(k^2)(g_{\mu \nu
}-\xi k_\mu k_\nu /k^2)$ by $(1-1/K(k^2)) D(k^2) (g_{\mu \nu }-k_\mu
k_\nu /k^2)$. This is transversal, therefore it is most easily
discussed in the Landau gauge $\xi =1$ where one simply gets a factor
$1-1/K(k^2) = \int d\mu ^2\bar{\kappa }(\mu ^2)/(k^2-\mu ^2+i\epsilon
) + O(1/k^4)$ which multiplies the original propagator. Obviously one
has a suppression by an additional power of two in the denominator,
therefore only insertions into the graphs (3),(4),(15) lead to
divergent contributions, all the others become convergent. In Landau
gauge we find an additional contribution of -4 to $l^{(3)}$, -4 to
$l^{(4)}$, and 12 to $l^{(15)}$.

An immediate result is, that now $\gamma _{ghost}$ becomes independent
of the spectral function $\kappa (m^2)$; because of the relation
$l_\sigma =-h_\sigma /(1+m^2)$ in (6.16) the square bracket in (6.20)
is proportional to $1-\int dm^2\kappa (m^2)/(1+m^2) = K$ which
cancels against the $1/K$ in front which now survives when $\delta
^2g_0^2/K$ is replaced by $g^2/K$ according to (6.24). This is
quite welcome, because the unpleasant linear contribution in $\kappa $
has thus disappeared. The same is, however, not true for $\gamma
_{gluon}$ and for $\beta $. This result will also hold in higher
orders, because higher order insertions make all graphs finite.

The additional contributions of order $\delta ^5g^3$ just discussed
(remember that we did not take into account terms of order $\delta
^5g^5$) all vanish if $\kappa (m^2)=0$. Comparing order $\delta ^3$
and $\delta ^5$, FAC would now trivially give the solution $\kappa
\equiv 0$ which is just what we expect. Inclusion of the two-loop
contributions $\sim \delta ^5g^5$ would shift this to a non-trivial
solution for $\kappa $ and finally result in a non-perturbative
solution for the $\beta $-function.

For PMS, on the other hand, there is really a problem. The
contributions $\sim \kappa $ are still there, and the optimized
$\delta $-expansion would not reproduce the lowest order perturbative
result!

\setcounter{equation}{0}\addtocounter{saveeqn}{1}%

\setcounter{equation}{0}\addtocounter{saveeqn}{1}%

\section{Conclusions }

At present we cannot offer a convincing explanation for the results
obtained above. Of course we cannot exclude the possibility of a
calculational error in our formulae, although we have carefully
checked them. Assuming that they are correct, it remains a puzzle why
PMS fails to reproduce the lowest order perturbative result. This is
particularly confusing because the puzzle persists to any order in
$\delta $ if $g$ is small. An encouraging result would have consisted
in a cancellation of all the terms linear in $\kappa (m^2)$ appearing
in the $\beta $-function, i.e. $l_\beta =0$. The quadratic terms would
then lead to an extremum at $\kappa (m^2) \equiv 0$, thus reproducing
the results of one-loop ordinary perturbation theory. In a two-loop
calculation, which might be feasible with more intensive computer
help, one would then expect a non-trivial solution for $\kappa (m^2)$
and a non-perturbative determination of the $\beta $-function. But
this was not what we obtained. On the other hand, FAC could lead to
interesting non-perturbative results in two-loop order which include
ordinary perturbation theory when expanded with respect to $g$.

Let us now discuss a different possible approach which was already
briefly mentioned in sect. 5. One could calculate the renormalized
gluon propagator in terms of the general spectral function $\bar
{\kappa }$ introduced in (5.1) and remove the remaining divergent
contributions by a definite prescription. This would be complementary
to - and of course much more complicated than - the calculations in
sect. 6 where we concentrated on the divergent terms. Finally one
could determine $\bar {\kappa }$ using FAC or PMS.

Note, however, that our ansatz in (5.1) implies $K(k^2)\rightarrow 1$
and, accordingly, $D(k^2) \rightarrow 1/k^2$ for $k^2\rightarrow
\infty $. The renormalized propagator, as determined by PMS, could
nevertheless show the correct asymptotic behavior as obtained from the
renormalization group, namely $\Delta (k^2) \sim [\ln k^2]^{-(10+3\xi
)/44}/k^2 $. (See e.g. \cite{mapa}). If one would apply FAC, on the
other hand, the bare and the renormalized propagator would be
identical by definition of the method. In this case one should either
use the special gauge $\xi =-10/3$ for which $\gamma _{gluon} = 0$ and
$\Delta (k^2)\sim 1/k^2$, or, more generally, write down a twice
subtracted dispersion relation for $K(k^2)$ with a finite subtraction
point. A calculation of this type, though complicated, appears
possible and will be undertaken in the future.

Finally we would like to mention a conceptional problem. The
connection between the bare and the renormalized coupling constant
will always start with $g\sim \delta g_0$. When $g_0$ is eliminated,
higher powers $(\delta g_0)^n$ become proportional to $g^n$, i.e.
become independent of $\delta $. So the clear bookkeeping of powers of
$\delta $ is somehow blurred by the renormalization procedure.

The present approach appears complicated. We interpret this as a
reflection of the fact that QCD is complicated and that
non-perturbative results can only be obtained with considerable
effort. We believe that interesting information can be extracted from
the general expressions presented here, and even more interesting
information from a two-loop calculation which might be feasible. The
puzzle in connection with PMS is not understood at present. Any
suggestions are welcome.\\[2ex]

{\large \bf Acknowledgement:} I thank N. Brambilla, M. Jamin, G.
Prosperi, I. Solovtsov, and A. Vairo for valuable discussions and I.
Bender for a careful reading of the manuscript.

\newpage
\begin{appendix}
\section{Appendix}

We give here some technicalities how the divergent contributions of
the string terms can be extracted. The problematic terms are those
where the internal momentum $k$ is multiplied with a string parameter
$s$ which has to be integrated from 0 to 1. For $s\approx 0$ there is
no suppression by a power of $k$ in the denominator, therefore a more
detailed analysis is necessary.

In a first step introduce the spectral representations (5.1) or (6.2),
(6.4) for $K$ and $D$, as well as the resulting representations for
$K^\mu $ and $K^{\mu \nu }$, and rotate to euclidean space as usual.
In all denominators without string parameters $s$ or $s'$ expand with
respect to $1/k^2$, i.e.

\begin{equation} \frac{1}{(k-q)^2+\mu ^2} =\frac{1}{k^2} +
\frac{2qk}{k^4} + \frac{4(qk)^2}{k^6} - \frac{q^2+\mu ^2}{k^4}+ \cdots
\end{equation}
up to the order where the further terms become ultraviolet finite.
The apparent infrared singularities arising from the expansion are
spurious.

In terms with only one string, e.g. (8), we have a further denominator
of the form $1/[(q-sk)^2+\mu ^2]^2$ arising from the spectral
representation of $K^\rho (q-sk)$. The substitution $k=k'/s$ makes the
integration over $k'$ finite. But for every power $k^{-n}$ in the
integrand a power $s^{n-d} = s^{n-4+2\epsilon }$ appears in
$d=4-2\epsilon $ dimensions. The $s$-integration gives a factor
$(n-3+2\epsilon )^{-1}$ and thus a divergent contribution if $n=3$.

Next we discuss the terms with two string integrations over $s,s'$.
Terms (10),(11) are finite, in (12),(13) substitute $s=r(1-t),s'=rt$.
This gives a factor $r$ from the Jacobian. Next substitute $k=k'/r$
and perform the integral over $r$ from 0 to 1 as before. This gives
again some divergent factors, the remaining integral is finite. The
second order term in (19) is finite, in (18) and (20) substitute
$t=s-s',t'=(s+s'-1)/2$. The $t'$-integration can be trivially
performed and gives a factor $(1-t)$, then substitute $k=k'/t$ as
before. After integration over $k'$ the various terms in (20)
cancel, so that this graph, contrary to what one would expect, does
not give a divergent contribution.

\end{appendix}

{\Large {\bf Figure Captions}}\\[2ex]

{\bf Fig. 1:} Interaction terms in momentum space.
All momenta are incoming. A thick line denotes the presence of a
factor $K$. Thick lines with gluons attached at the interior of the
lines arise from the expansion of the string $U(x,y)$. They are
associated with a factor $K^\mu = \partial ^\mu K$ or $K^{\mu \nu }=
\partial ^\mu \partial ^\nu K$, respectively. String parameters $s,s'$
are integrated from 0 to 1.\\[2ex]

{\bf Fig. 2:} Contributions to the gluon vacuum polarization. \\[2ex]

{\bf Fig. 3:} Ghost self-energy. \\[2ex]

{\bf Fig. 4:} Contributions to the ghost gluon vertex.

\includegraphics{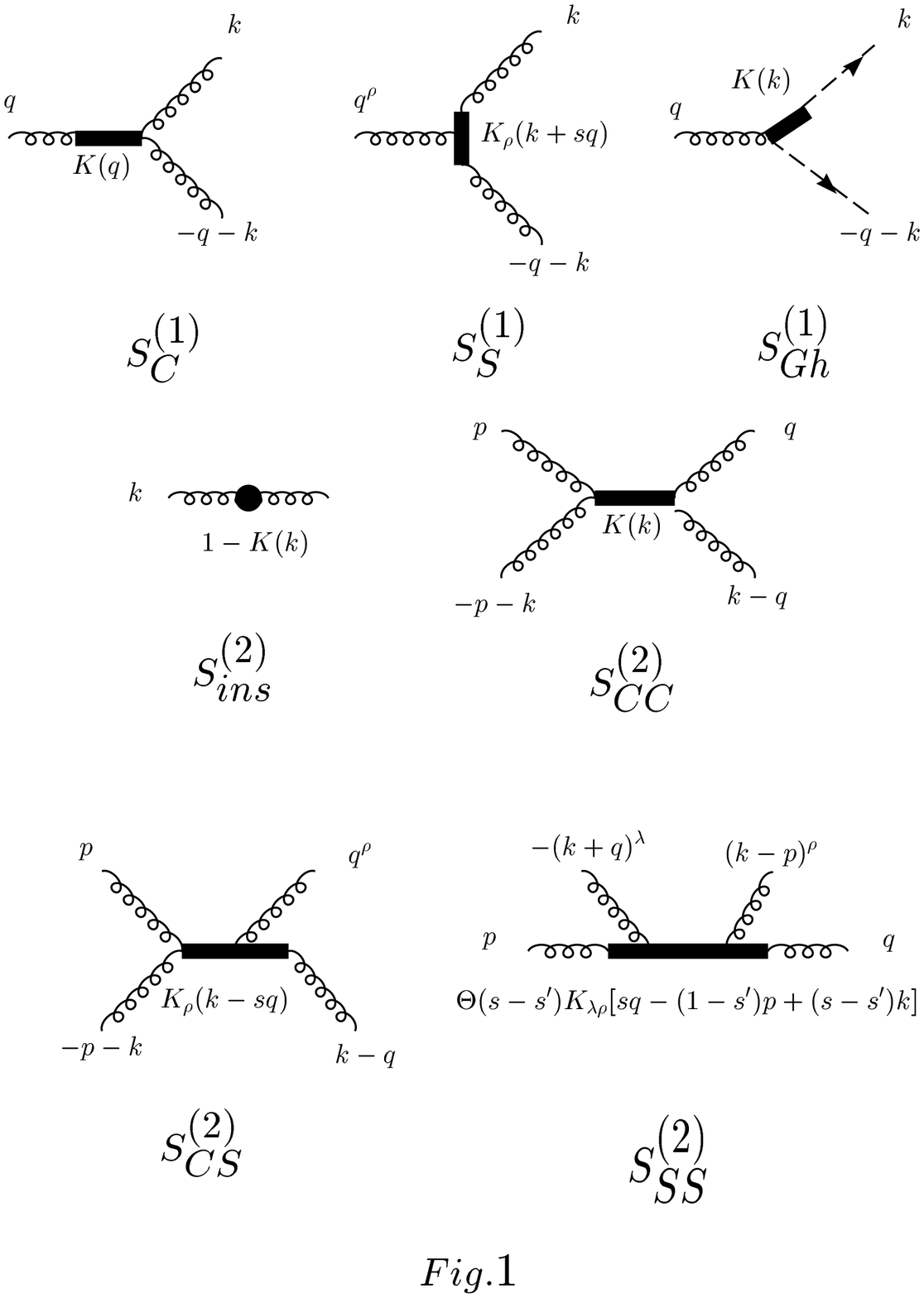}
\includegraphics{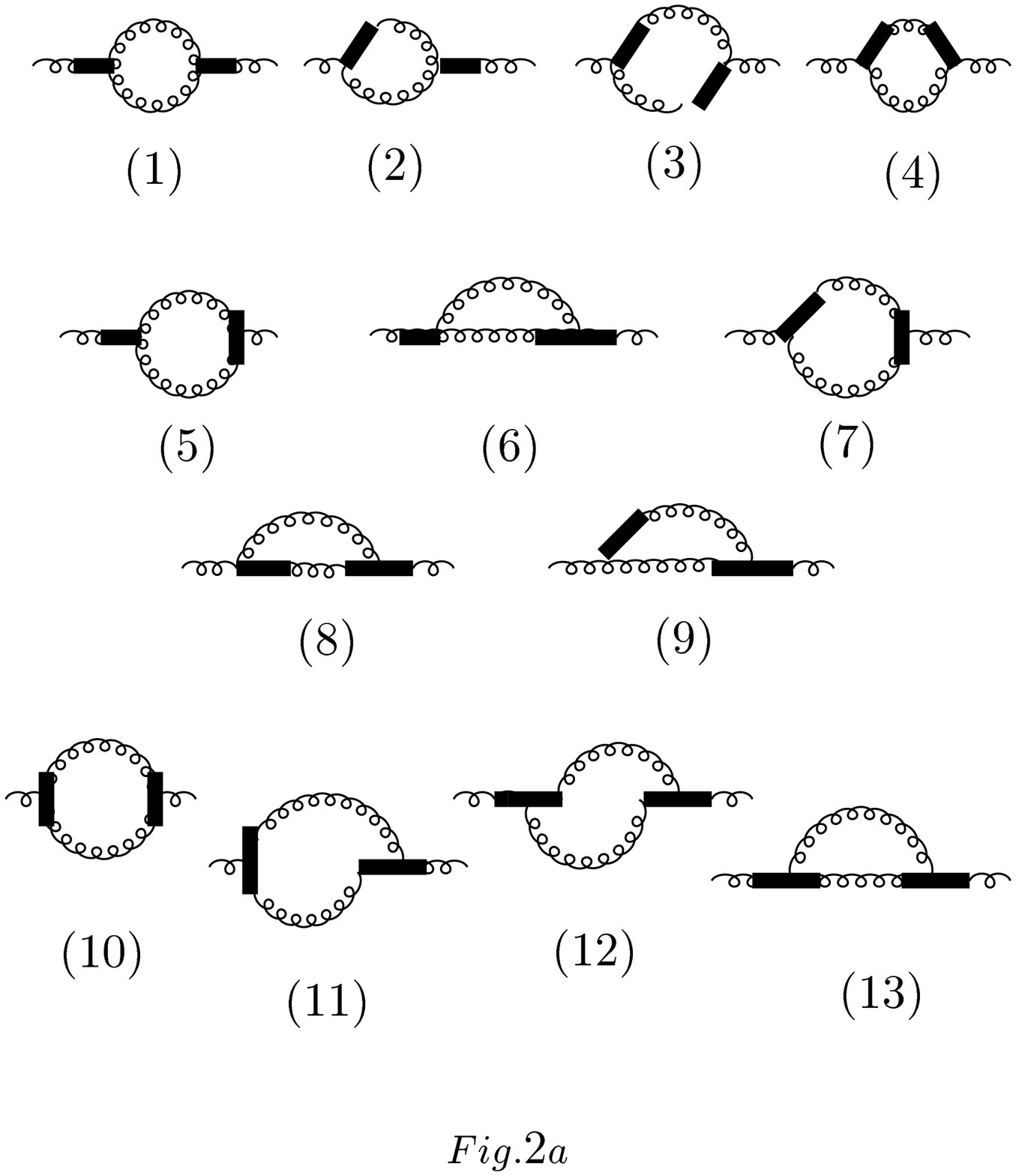}
\includegraphics{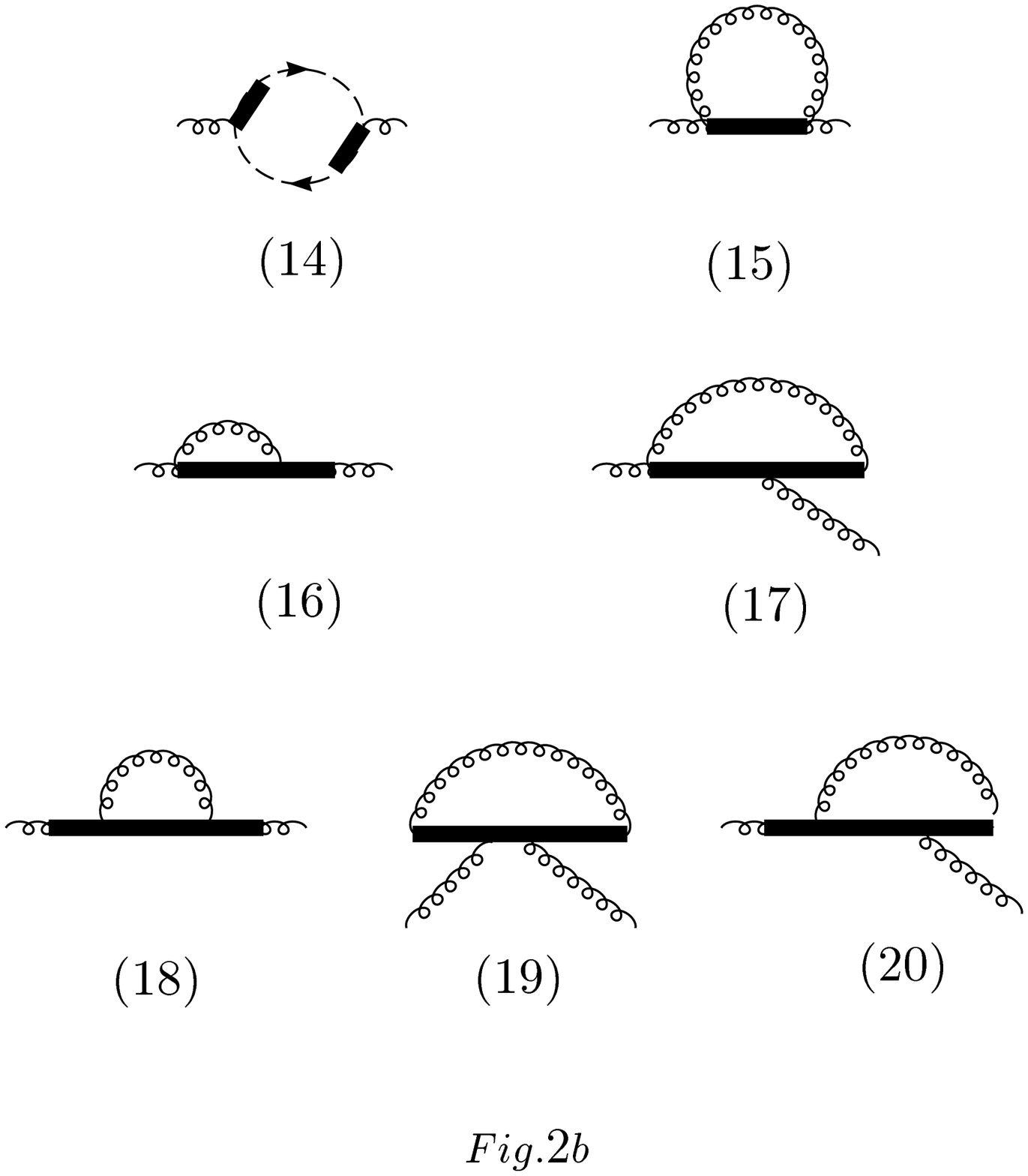}
\includegraphics{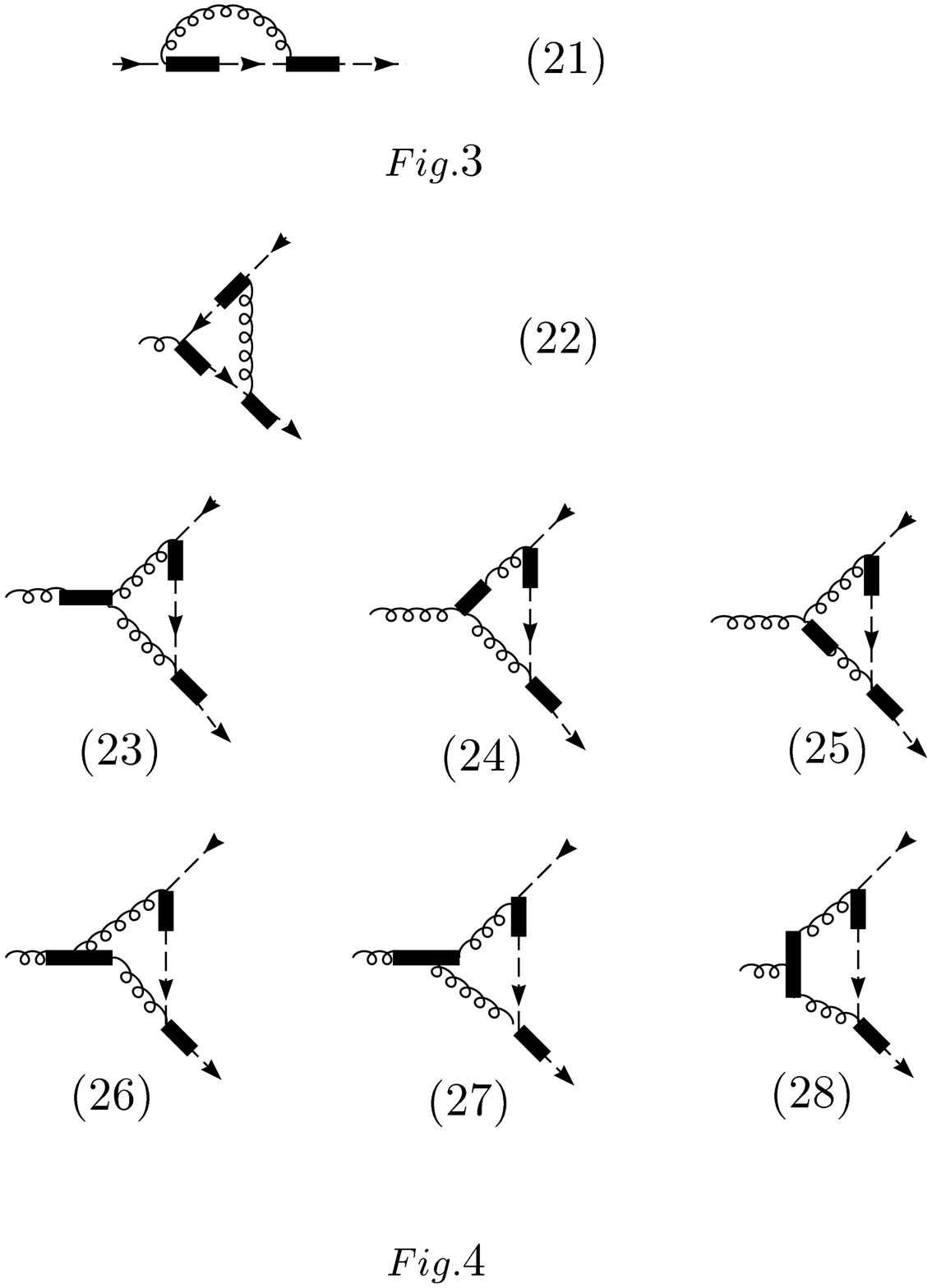}

\end{document}